\documentclass{ws-mplb}
\usepackage[none]{hyphenat}
\begin{document}


%
\catchline{}{}{}{}{}
%

\title{Phase diagram of one-dimensional earth-alkaline cold
fermionic atoms
}

\author{\footnotesize H. Nonne}

\address{Laboratoire de Physique Th\'eorique et
Mod\'elisation, CNRS UMR 8089,
Universit\'e de Cergy-Pontoise, Site de Saint-Martin,
F-95300 Cergy-Pontoise Cedex, France}

\author{\footnotesize E. Boulat}

\address{Laboratoire Mat\'eriaux et Ph\'enom\`enes Quantiques,
Universit\'e Paris  Diderot, 2 Place Jussieu,
75205 Paris Cedex 13, France}

\author{\footnotesize S. Capponi}

\address{Laboratoire de Physique Th{\'e}orique, Universit{\'e} de Toulouse,
UPS (IRSAMC), F-31062 Toulouse, France}

\author{\footnotesize P. Lecheminant}

\address{Laboratoire de Physique Th\'eorique et
Mod\'elisation, CNRS UMR 8089,
Universit\'e de Cergy-Pontoise, Site de Saint-Martin,
F-95300 Cergy-Pontoise Cedex, France}

\maketitle


\begin{abstract}
The phase diagram of one-dimensional earth-alkaline fermionic
atoms and ytterbium $171$ atoms
is investigated by means of a low-energy approach
and  density-matrix renormalization group calculations.
For incommensurate filling, four gapless phases with
a spin gap are found and consist of two superconducting
instabilities and two coexisting bond and charge density-waves instabilities. In the half-filled case, seven Mott-insulating
phases arise with the emergence of four non-degenerate
phases with exotic hidden orderings.

\end{abstract}


The recent experimental progresses achieved in trapped
ultracold atomic gases augur great opportunities to explore
the physics of strong correlations in clean systems and
within a wide range of parameters, owing to
optical lattices and the high tunability  of Feshbach
resonances. The attention of the cold atoms community has
recently been called to fermionic earth-alkaline atoms
such as strontium (Sr) and to atoms endowed with a similar
electron structure, such as ytterbium (Yb), which
have been cooled down to reach the quantum degeneracy~\cite{Desalvo,fukuhara}.
The interest in those atoms stems from the presence of a ground
state $^1S_0$ (``$g$'') and a metastable excited state
$^3P_0$ (``$e$'') between which transitions are forbidden;
moreover, both states have zero electronic angular momentum
$J$, so that the nuclear spin $I$ is decoupled from the
electronic spin~\cite{Gorshkov2009}. The scattering lengths
are independent of the nuclear spins and this results in systems
with an extended SU($N=2I+1$) symmetry that can be realized
without any fine-tunning; $N$ can be as large as 10 with
$^{87}$Sr ($I=9/2$).
In this respect, earth-alkaline cold atoms seem to be very
promising for the experimental realization of high-symmetry systems,
and the investigation of exotic many-body physics~\cite{Gorshkov2010,cazalilla,Xu2010,Wu2006}.

In this paper, we will focus on the one-dimensional (1D)
$N=2$ case, that is on systems of earth-alkaline like atoms
with a nuclear spin $I=1/2$ such as $^{171}$Yb atoms or earth-alkaline
atoms where only two nuclear spin states have been trapped.
The general $N$ case will be investigated elsewhere.
Besides, in addition to the U($1$)$_c$ charge
symmetry due to the conservation of the total number of
atoms, there is also a U($1$)$_o$
orbital symmetry, as a result of the conservation
of the number of atoms in each state ($g$ and $e$).
These systems display thus an SU($2$)$_s \times$U($1$)$_c\times$U($1$)$_o$
continuous symmetry. Moreover, for the sake of simplicity, we assume
an extra orbital Z$_2$ symmetry between the two
orbital levels $e$ and $g$ so that they play a similar role.
In that case, the effective
Hamiltonian model describing such systems
loaded in 1D optical lattice is:~\cite{Gorshkov2010}
\begin{eqnarray}
  {\cal H}&=& -t \sum_{i,l\alpha}[c^{\dagger}_{l \alpha,i}
  c^{\phantom{\dagger}}_{l \alpha,i+1}+{\rm H.c.}]
- \mu \sum_{i,l} n_{l,i}
+\frac{U}{2} \sum_{i,l} n_{l,i} \left(n_{l,i} - 1 \right) \nonumber\\
&&\qquad+V\sum_i n_{g,i}n_{e,i}
+V_{\rm ex}\sum_{i,\alpha,\beta} c_{g\alpha,i}^\dagger
c_{e\beta,i}^\dagger
c^{\phantom\dagger}_{g\beta,i}c^{\phantom\dagger}_{e\alpha,i},\label{modelalka}
\end{eqnarray}
where $\alpha,\beta=\uparrow,\downarrow$ is the SU($2$) spin
index and $l=g,e$ represents the two orbital levels.
In Eq. (\ref{modelalka}),
$c^{\dagger}_{l\alpha,i}$ are the four-components
fermionic creation operators on site $i$, and
$n_{l,i} = \sum_{\alpha} c_{l\alpha,i}^\dagger c^{\phantom\dagger}_{l\alpha,i}$ denotes
the density operator on orbital level $l=e,g$ and
site $i$.
We intend to investigate the
zero-temperature phase diagram of model (\ref{modelalka})
for incommensurate filling and
at half-filling. For this purpose, we will follow a
weak-coupling approach, supported by numerical simulations for moderate couplings, using
the Density Matrix Renormalization Group (DMRG) algorithm~\cite{DMRG}.\\

\section{Incommensurate filling}

Since we are interested in the low-energy properties of
model (\ref{modelalka}), the starting point of our approach
lies in the weak-coupling regime $|U,V,V_{\rm ex}|\ll t$.
In this regime, only the long-distance properties of the
model matter and we perform a continuum limit
by linearizing the dispersion relation for
non-interacting fermions about the two Fermi points ($\pm
k_F$). At incommensurate fillings, there is a spin-charge
separation: the charge and the spin, orbital degrees of freedom are
decoupled and can be treated separately. The charge degrees
of freedom are gapless and exhibit the characteristics of
the Luttinger liquid~\cite{book}.
The low-energy properties of the
remaining spin and orbital degrees of freedom can
be described in terms of six Majorana (real) fermions $\xi^a_{R,L}$
($a=1, \ldots, 6$) as in the spin-orbital ladder~\cite{azaria}.
The interacting part of the
corresponding effective
Hamiltonian density reads as follows:
\begin{eqnarray}
  \mathcal{H}^{\rm int}_{so}&=& \frac{g_1}{2}\left(\sum_{a=1}^3
  \xi_R^a\xi_L^a\right)^2+g_2\left(\sum_{a=1}^3
  \xi_R^a\xi_L^a\right)\left(\sum_{a=4}^5
  \xi_R^a\xi_L^a\right)\nonumber\\
  &&\qquad\qquad+\xi_R^6\xi_L^6\left[g_3\sum_{a=1}^3
  \xi_R^a\xi_L^a+g_4\sum_{a=4}^5 \xi_R^a
  \xi_L^a\right] +\frac{g_5}{2}\left(
  \sum_{a=4}^5 \xi_R^a\xi_L^a\right)^2,
  \label{Hsincommensurate}
\end{eqnarray}
with $g_1=-a_0(U+V_{\rm ex})$, $g_2=-a_0 V$,
$g_3=-a_0(U-V_{\rm ex})$, $g_4=-a_0 (V-2V_{\rm ex})$,
and $g_5=a_0(U-2V+V_{\rm ex})$, $a_0$ being the lattice spacing.
The one-loop renormalization group (RG) equations
for that model can be found in Ref.~\protect\refcite{BoulatdualitiesNPB2009}.
The RG analysis leads to the emergence of
four different spin-gapped phases.
Two phases have coexisting  bond
and density wave instabilities and can be distinguished
by their properties under the Z$_2$ exchange
symmetry $(e \leftrightarrow g)$.
Spin-Peierls (SP) and charge-density wave (CDW) instabilities
are even under Z$_2$
while SP$_\pi$ and
CDW$_\pi$ - or orbital-density wave (ODW) -
instabilities are odd. Their order parameters are:
\begin{eqnarray}
{\cal O}_{i}^{\rm CDW}=e^{-2 i k_F  a_0}
\sum_{l,\alpha} c_{l \alpha,i}^\dag c_{l \alpha,i},
&\,&{\cal O}_{i}^{\rm CDW_\pi}=
e^{-2 i k_F a_0} \sum_{l, \alpha}
\epsilon_l c_{l \alpha,i}^\dag
c_{l \alpha, i},\nonumber\\
{\cal O}_{i}^{\rm SP}=e^{-2 i k_F  a_0}
 \sum_{l,\alpha} c_{l \alpha,i}^\dag
 c_{l \alpha,i+1},
&\,&{\cal O}_{i}^{\rm SP_\pi}=e^{-2 i k_F a_0}
\sum_{l,\alpha} \epsilon_l c_{l \alpha,i}^\dag
c_{l \alpha,i+1},\label{orderdegincom}
\end{eqnarray}
with $\epsilon_l=\pm 1$ for $l=g,e$.
The SP instability has dimerised pairs of atoms in
the $g$ or in the $e$ states; the CDW instability
has alternating fully occupied and empty states,
and the CDW$_\pi$ (or ODW) instability is an
alternation of sites with two atoms in the $g$ state
and sites with two atoms in the $e$ state.
Figure~\ref{fig:degeneratephases} (next section)
shows a pictural representation of these instabilities.

\begin{figure}[!h]
\includegraphics[width=0.49\linewidth,clip]{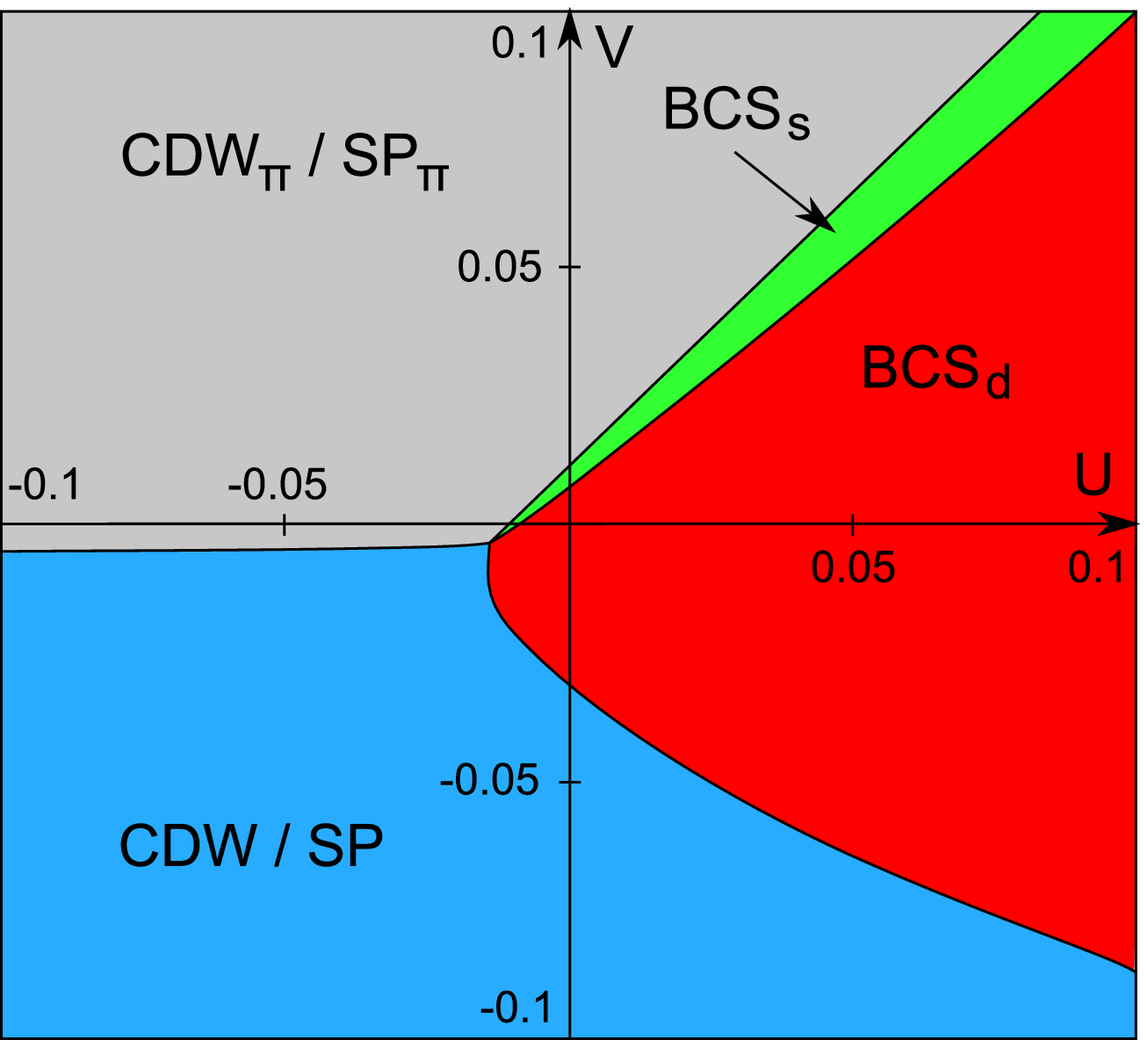}
\includegraphics[width=0.49\linewidth,clip]{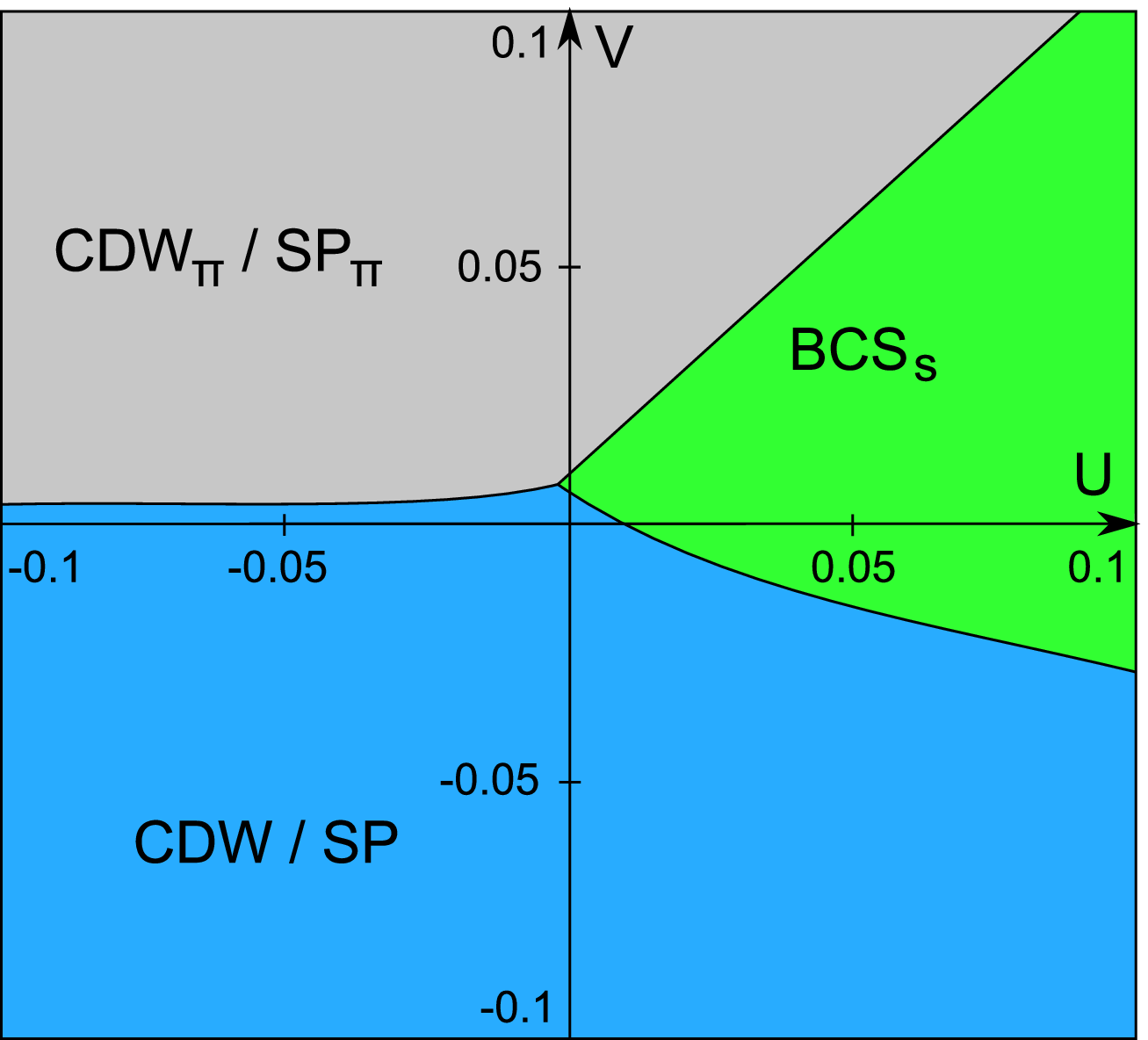}
\caption{Phase diagram of model (\ref{modelalka}) at incommensurate filling for $V_{\rm ex}=-0.01t$ (left) and $V_{\rm ex}=0.01t$ (right).}
\label{fig:phasediagincom}
\end{figure}

The two other phases
are superconducting instabilities
and differ in the Z$_2$ exchange symmetry of
their pairing operators:
\begin{eqnarray}
&&{\cal O}^{\rm BCSs}_{i}= c_{g\uparrow,i+1}c_{e\downarrow,i}
-c_{g\downarrow,i+1}c_{e\uparrow,i}
- (e \leftrightarrow g),\nonumber\\
&&{\cal O}^{BCSd}_{i}= c_{g\uparrow,i}c_{e\downarrow,i}
-c_{g\downarrow,i}c_{e\uparrow,i}.\label{orderBCSincom}
\end{eqnarray}
The zero-temperature phase diagram is sketched in
Fig.~\ref{fig:phasediagincom}
depending on the sign of $V_{\rm ex}$.

\section{Half-filling case}
\subsection{Weak-coupling approach}
\begin{figure}[!h]
\includegraphics[width=0.49\linewidth,clip]{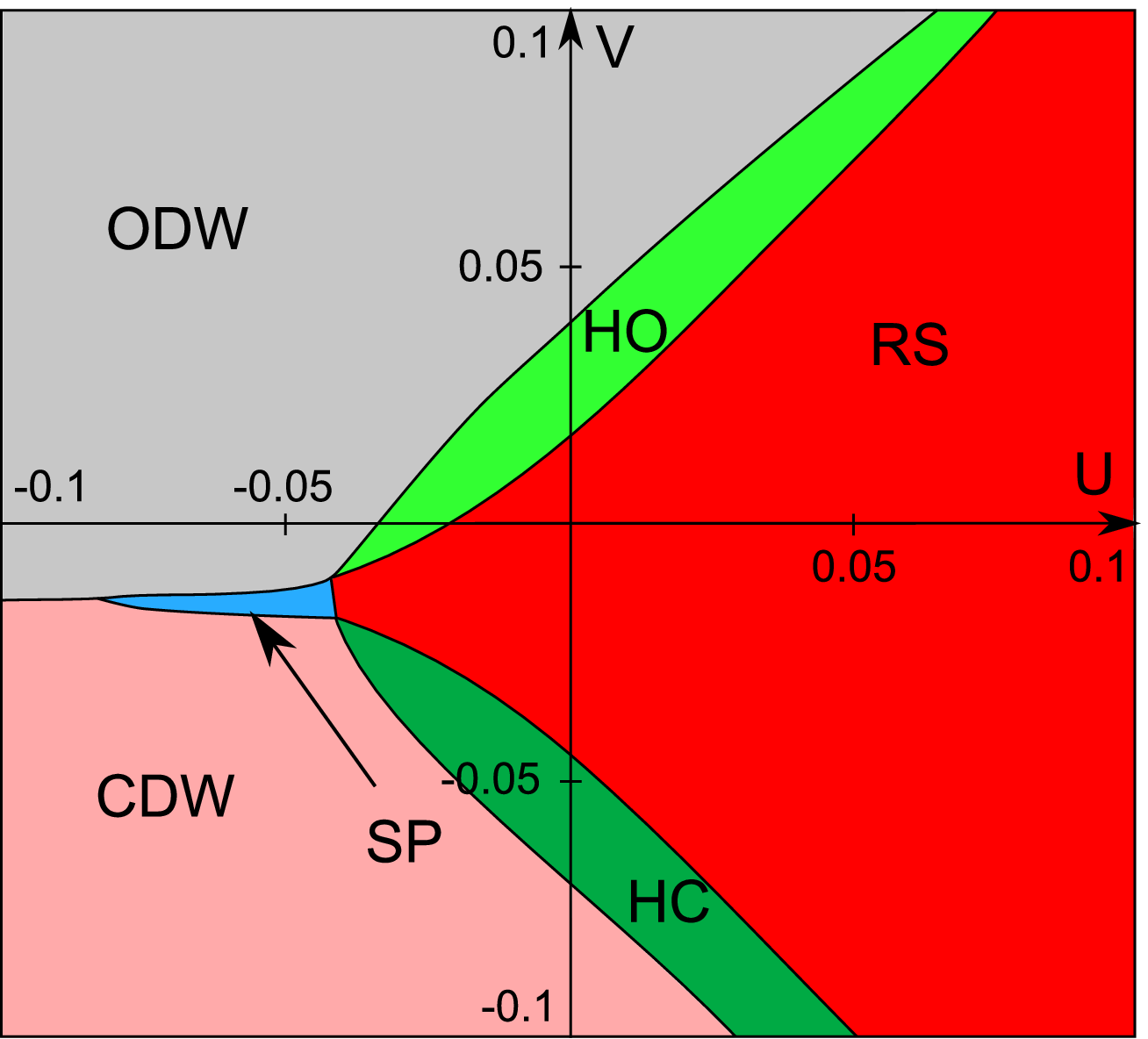}
\includegraphics[width=0.49\linewidth,clip]{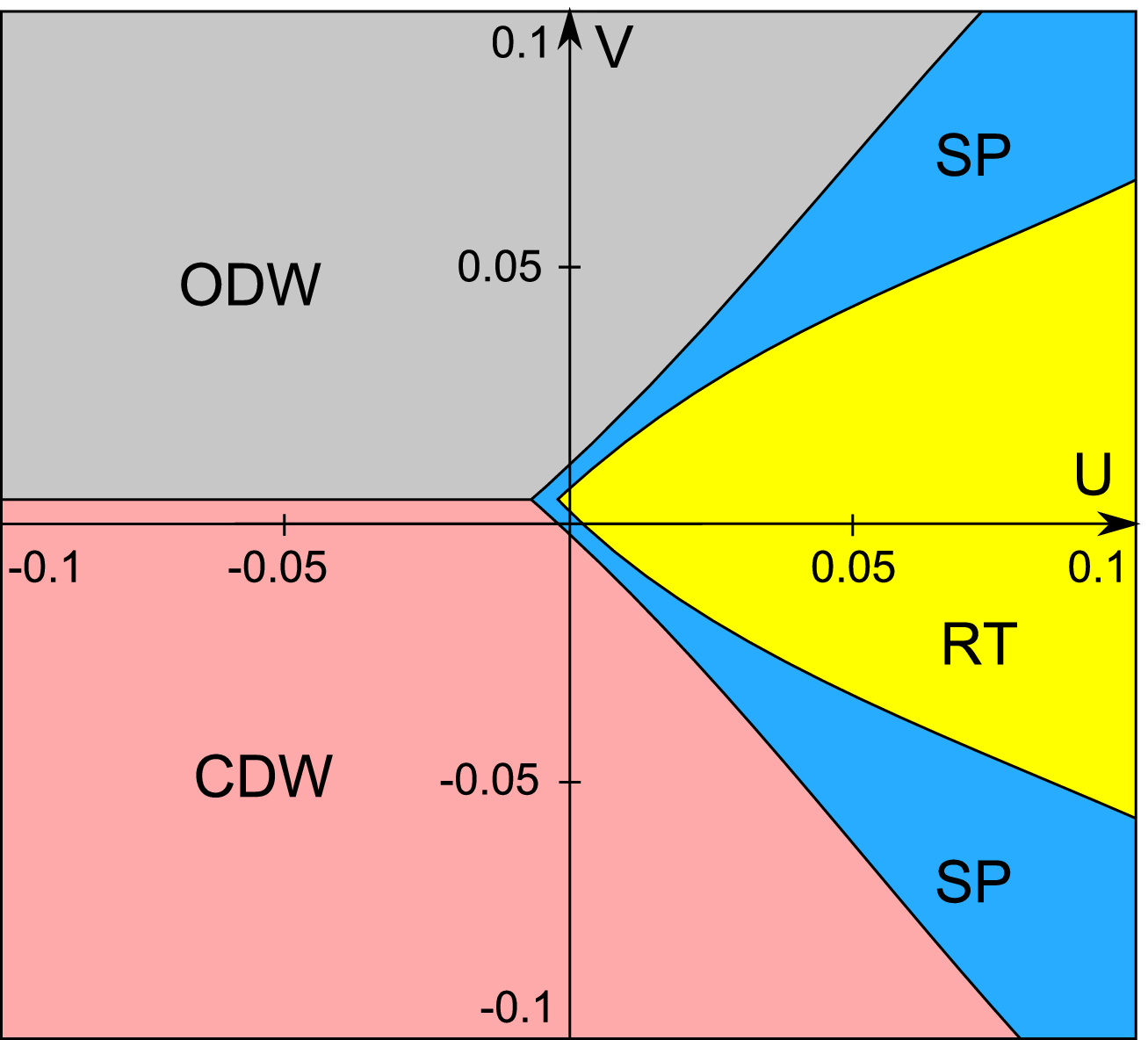}
\caption{Phase diagram of model (\ref{modelalka}) at
half-filling for $V_{\rm ex}=-0.03t$ (left) and $V_{\rm ex}=0.01t$ (right).}
\label{fig:phasediaghalff}
\end{figure}

At half-filling, there is no spin-charge separation
anymore since umklapp processes
couple these degrees of freedom in sharp contrast
to the half-filled Hubbard chain~\cite{book}.
The interplay between charge and spin-orbital degrees of
freedom can be investigated by introducing
two additional Majorana fermions $\xi_{R,L}^{7,8}$ for the charge
sector. The umklapp processes generate new terms which add
to the effective interacting Hamiltonian
(\ref{Hsincommensurate}):
\begin{eqnarray}
  {\cal H}_{\rm umklapp}&=&
  \frac{g_6}{2}\left(\sum_{a=7}^8\xi_R^a\xi_L^a\right)^2
  \nonumber\\
  && + \left(
  \xi_R^7\xi_L^7+\xi_R^8\xi_L^8\right)
  \left[g_7\sum_{a=1}^3 \xi_R^a
  \xi_L^a+g_8\sum_{a=4}^5 \xi_R^a\xi_L^a
  +g_9\xi_R^6\xi_L^6\right], \label{umklapp}
\end{eqnarray}
where $g_6=a_0(U+2V-V_{\rm ex})$, $g_7=a_0(V-V_{\rm ex})$,
$g_8=a_0 U$, and $g_9=a_0(V+V_{\rm ex})$.
The analysis of the one-loop RG equations for the
Majorana models (\ref{Hsincommensurate}, \ref{umklapp})
is detailed in Ref.~\cite{NonneGeneralizedHundmodel2010}.
The resulting zero-temperature phase diagram is depicted in Fig.
\ref{fig:phasediaghalff},

\begin{figure}[!h]
\centering
\includegraphics[width=1\linewidth]{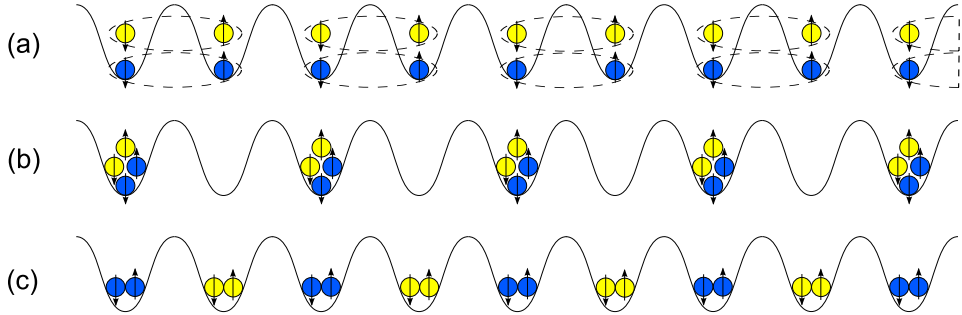}
\caption{(color online) The blue atoms are in the
ground state, and the yellow color represents
the excited state. (a) The SP phase contains dimerised
pairs of atoms in the $g$ or in the $e$ states;
(b) the CDW phase swings between fully occupied and empty states;
(c) the ODW is an alternation of sites with two atoms
in the $g$ state and sites with two atoms in the $e$ state.}
\label{fig:degeneratephases}
\end{figure}

Seven Mott-insulating phases have been found in the phase
diagram, all fully gapped in charge, orbital, and spin
degrees of freedom; three of them are two-fold
degenerate and four are non-degenerate. As regards
the three degenerate phases, depicted in Fig.
\ref{fig:degeneratephases}, their order parameters
were given in Eq. (\ref{orderdegincom}):
one is the SP phase (${\cal O}_{i}^{\rm SP}$), a
second one is the CDW phase
(${\cal O}_{i}^{\rm CDW}$), and the third one
is the ODW phase
(${\cal O}_{i}^{\rm ODW}={\cal O}_{i}^{\rm CDW_\pi}$).
In the incommensurate case, these orders have a power-law
decay while true long-range ordering occurs in the
half-filled case.

\begin{figure}
\centering
\includegraphics[width=1\linewidth]{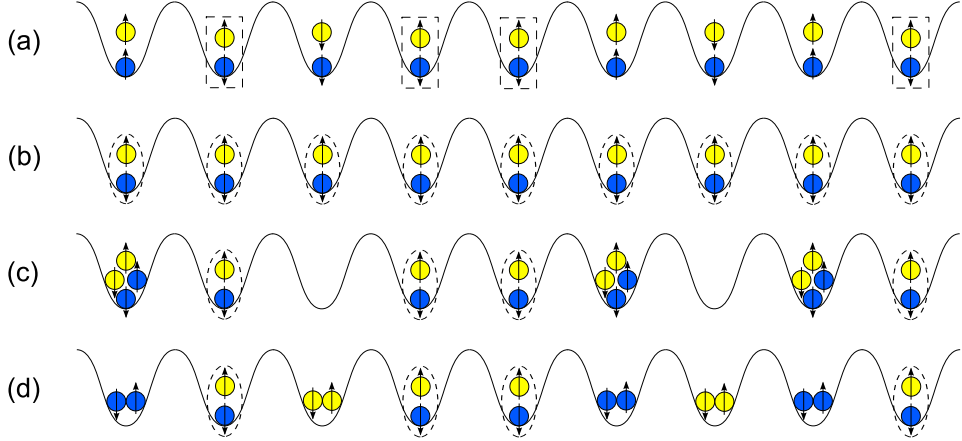}
\caption{(color online) The blue atoms are
in the ground state, and the yellow color
represents the excited state. (a): The RT
phase displays alternating $S^z=\pm 1$ states
with an arbitrary number of $S^z=0$
(spin triplet) states in between;
(b) the RS phase consists in spin singlets formed
on each site by two atoms, one in the $g$ state
and the other in the $e$ state.
(c) the HC phase shows alternating $n_i=4,0$ states
with an arbitrary number of singlets in between;
(d) the HO phase exhibits alternating
$(n_g-n_e)/2=\pm 1$ states with again an arbitrary
number of singlets in between.}
\label{fig:nondegeneratephases}
\end{figure}

The first non-degenerate phase is the rung triplet
(RT) phase, so called since it is related to the RT
phase of the two-leg spin-1/2 ladder, with the
orbital states $g,e$ playing the role of the legs.
This gapped phase is adiabatically connected to the
Haldane phase of the spin-1
Heisenberg chain~\cite{book} and displays a
dilute order which consists in an alternation of
$S^z=S^z_g+S^z_e=+1$ (with $S^z_l=[c_{l\uparrow,i}^\dagger c^{\phantom\dagger}_{l\uparrow,i} -c_{l\downarrow,i}^\dagger c^{\phantom\dagger}_{l\downarrow,i}]/2$) and $S^z=-1$ states with an
arbitrary number of $S^z=0$ states in between
(see Fig. \ref{fig:nondegeneratephases} (a)).
The second non-degenerate phase is a rung singlet
(RS) phase where the two orbital states form a spin
singlet state on each site (Fig.
\ref{fig:nondegeneratephases} (b)).
The third phase is called Haldane charge (HC) phase
since it belongs to the same type as the Haldane
(spin) phase above, but now, the correct quantum
number to characterize the dilute order is the
occupation number $n_i$. States with $n_i=4$ and
$n_i=0$ alternate, and there is an arbitrary number
of singlet ($n_i=2$) states in between. It is easily
seen (Fig. \ref{fig:nondegeneratephases} (c)) that
HC interpolates between the CDW and the RS phase.
This HC phase has been first identified in the context
of 1D spin-3/2 cold alkaline fermions~\cite{Nonnehaldanecharge32}.
Finally, the last non-degenerate phase is called
Haldane orbital (HO) and the alternation takes
place between states with $(n_g-n_e)/2=\pm 1$ with
again an arbitrary number of spin singlets
$(n_g-n_e)/2=0$ in between (Fig.
\ref{fig:nondegeneratephases} (d)). This phase
interpolates between the ODW and the RS phases.

\subsection{Numerical approach}

We now carry out numerical calculations using DMRG in order to
investigate the various phase diagrams, and we focus on the half-filled case.
For a finite system
of  size $L$, we can fix several quantum numbers: the total number of particles ${Q_c=\sum_{i,l} n_{l,i}=2L}$, the $z$-component of the total spin of the whole system
${S^z_{\rm tot}= \sum_{i} (S^z_{e,i}+S^z_{g,i})}$, as well as
the population difference between the two orbital levels $Q_o = \sum_i (n_{e,i}-n_{g,i})$. Typically,
we keep up to 1600 states, which allow to
have an error below $10^{-6}$, and we use open boundary
conditions (OBC).

The low-energy approach predicts seven insulating phases in the phase diagram. In order to identify the degenerate Mott phases (CDW, ODW, and SP), we compute
local quantities (bond kinetic energy, local density, etc.). In order
to detect non-degenerate Mott phases, we investigate the
presence or absence of various
edge states with quantum numbers $(Q_c,S^z_{\rm tot},Q_o)$,
respectively equal to $(L,1,0)$, $(L+2,0,0)$, and $(L,0,1)$ for the
RT, HC, and HO phases. The RS phase is identified as
being non-degenerate and does not exhibit any kind of edge states with OBC.

\begin{figure}[!h]
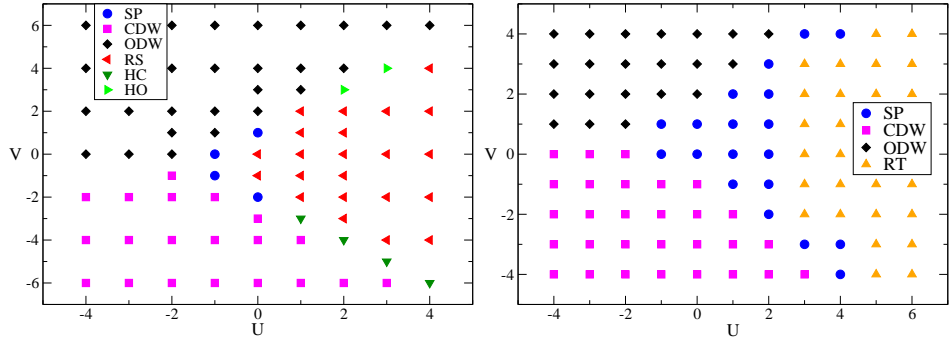

\includegraphics[width=0.49\linewidth,clip]{phase_diag_Yb_Vgex=-1}
\includegraphics[width=0.49\linewidth,clip]{phase_diag_Yb_Vgex=1}
\caption{Numerical phase diagram of model (\ref{modelalka}) at
half-filling for $V_{\rm ex}=-t$ (left) and $V_{\rm ex}=t$ (right). We fix $t=1$ as unit of energy.}
\label{fig:numphasediaghalf}
\end{figure}

The resulting zero-temperature phase diagrams are shown in
Fig.~\ref{fig:numphasediaghalf} for two choices of $V_{\rm ex}$.
The overall topology of these two phase diagrams is in agreement
with the low-energy results plotted in Fig.~\ref{fig:phasediaghalff}
since we confirm the presence of the seven insulating phases. There is a fair concordance for positive $V_{\rm ex}$. However, there is a little
discrepancy between the two approaches for negative $V_{\rm ex}$: the small SP, HC and HO pockets do not match exactly. Still, we emphasize that the weak coupling phase diagram was obtained for much smaller values of $|V_{\rm ex}|$; furthermore, the weak coupling approach shows that the SP pocket shifts and then quickly disappears for increasing $|V_{\rm ex}|$, in favor of the HC and HO phases. This indicates that the position of the SP pocket is not fully significant and strongly depends on the value of $V_{\rm ex}$.

\section{Concluding remarks}
We have investigated the phase diagram of ultracold 1D
fermionic systems
of earth-alkaline like atoms with nuclear spin ($I=1/2$),
at incommensurate filling and at half-filling. We showed that
at incommensurate filling, while the charge degrees of
freedoms remain gapless, the spin and orbital degrees of freedom
are all gapped. The phase diagram displays four different
phases with coexisting SP and CDW or with BCS instabilities.\\
At half-filling, seven Mott-insulating phases arise.
Three of them are two-fold degenerate (SP, CDW, and ODW
phases) and four are non-degenerate (RT, RS, HC, and HO phases).
Among the non-degenerate phases, three are of the
Haldane type (related to the physics of spin chains)
and show hidden dilute order.
We hope that, in
the light of the recent experiments of
earth-alkaline-atoms and $^{171}$Yb atoms, future experiments will
disclose part of the richness of the phase diagram
presented in this paper.


\end{document}